# Compact Non-Volatile Multilevel Sb$_2$Se$_3$ Electro-Optical Switching in the Mid-Infrared Group-IV-Photonics Platform

Richard Soref [a], Francesco De Leonardis [b], Martino De Carlo [b], Vittorio M. N. Passaro [b,*]

[a] *Department of Engineering, The University of Massachusetts, Boston, Massachusetts, 02125 US*
[b] *Dipartimento di Ingegneria Elettrica e dell'Informazione, Politecnico di Bari Via Edoardo Orabona n. 4, 70125 Bari, Italy*
*\*vittorio.passaro@poliba.it*

**Abstract**

This theoretical modeling and simulation paper presents designs and projected performances of two non-volatile, broadband, on-chip 2×2 electro-optical switches based upon the germanium-on-insulator (GeOI) photonic-electronic platform operating at the 2.5 μm mid-infrared wavelength. These compact devices facilitate large-scale integration on a "monolithic wafer" where all components are made of group-IV semiconductors. The switches are the two-waveguide directional coupler (DC) and the Mach-Zehnder interferometer (MZI). A thin-film graphene Joule-effect micro-heater is assumed on the planarized GeOI device to change the phase (reversably) of DC-slot-embedded Sb$_2$Se$_3$ phase-change material (PCM) from crystalline to amorphous. The MZI has this PCM within its slotted-arm waveguides. Simulations show high-performance bistable or multistable cross-bar switching in both devices. The 2×2 DC has an active coupling length of 17 μm, 130 nm gap, and a footprint of 5 μm x 31 μm. The device bandwidth is 30 nm over a wavelength range where cross and bar insertion losses IL are less than 0.3 dB, and where optical crosstalk is less than -15 dB. Results for the 2×2 MZI show crossbar switching attained with a 7.8 μm-length Sb$_2$Se$_3$ slot and a 5 μm x 51 μm switch footprint. Stable, multi-level switching in both devices is attained via partial amorphization. Thermal modeling shows that careful control of the voltage-pulse amplitude $V$ applied to graphene (rectangular pulse duration of 500 ns) can give 32 levels, for example, using $V$ in the range from 6.18 to 7.75 Volts.

## 1. Introduction

This theoretical paper analyzes the performance of 2×2 self-holding electro-optical switches enabled by optical phase-change material (PCM) embedded within the Ge waveguides of the germanium-on-insulator (GeOI) platform. The GeOI wafer is a relatively unexplored integrated-photonics wafer that is foundry-manufactured upon a large-area SiO$_2$-coated Si substrate, and the Ge strip-guides provide high transmission (low propagation loss) over the GeOI wavelength-of-operation range that extends from 1.6 μm out to 14 μm. Therefore, GeOI offers opto-electronics in the short-wave, mid-wave, and long-wave infrared (SWIR, MWIR, and LWIR). The present work focuses on the 2.5 μm wavelength.

The present paper presents numerical simulations and predictions about two broadband non-volatile electro-optical routing switches having very small footprint that facilitates large-scale integration (LSI) of devices for key OE chip applications such as neuromorphic computing, digital computing, optical sensing, optical network interconnects and programmable PICs.

Looking for a moment at the larger context of the GeOI wafer, it is possible in principle to use only group IV semiconductors, such as the SiGeSn alloys, to construct every active and passive component on the wafer: components in the photonic, electrooptic (EO), and electronic categories. This manufacturing will then be monolithic construction that is generally simpler and less expensive than on-wafer heterogeneous integration of devices made from materials that are "outside of group IV." Thus, GeOI exemplifies the monolithic "vision."

That vision includes a "complete suite" of group IV components on the wafer. Because extensive R&D has been done, the photonics community already has available a portion of the suite--the needed monolithic components of group IV photodetectors, volatile switches and 1 x 1 EO modulators. However, three group-IV devices are currently lacking within the suite; specifically, the 300K on-chip group IV laser diodes and optical amplifiers, and the bistable-or-multistable N x M EO routing switches described in this paper. We note that several laboratories are currently engaged in increasing the present 77K operating temperature of GeSn multi-quantum-well laser diodes up to 300K, a room-temperature result expected in the not-too-distant future. The switching wavelength selected here is the same as the emission wavelength of the GeSn laser diode and it is the detection wavelength of associated GeSn photodiodes.

Small-footprint devices are a contribution of this paper. The paper has the following sections: Section II is a background discussion of the prior art; while Section III considers innovative broadband 2×2 directional coupler (DC) crossbar switches where PCM fills the coupling zone. Detailed analysis is given. Section IV investigates the broadband Mach Zehnder Interferometer (MZI) 2×2 crossbar switches where a slot filled with PCM is formed in each arm. Performance metrics are studied. Section V discusses briefly GeOI PCM micro-resonator modulators and switches and photonic memories. Numerical simulation results and optimizations of the 2×2 DC switch are given in Section VI. Stable, multi-level operation of the MZI switch, related to partial amorphization, is investigated in Section VII, while Section VIII presents thermal analysis of temperatures reached in the DC PCM volume, as related to voltage-applied graphene. The MZI-related Section IX investigates the temperature distribution within the PCM slot versus voltage-on-graphene and relates that to amorphization. Finally, Section X summarizes the conclusions.

## 2. Materials and methods

### 2.1 Background discussion

In the prior art of volatile EO switches, where the mechanisms of free carrier, thermo-optic, ferroelectric-Pockels and Franz-Keldysh were deployed, the resonant devices (mainly micro rings) have indeed attained the small area that is wanted for the 2×2 devices. However, for the case of broadband switches, the attainment of a small $W_T \times L$ area (where $W_T$ and $L$ are the total width and length of the device, respectively, see Fig. 1) for those switches has proved elusive because $L$ is generally in the 0.2 to 1.0 mm range. That is because the electrically induced change of real index-of-refraction within the switching region has been less than, or much less than, 0.01.

PCMs are seen as an answer to this switch length problem because of the much larger (and reversible) change-in-index that PCMs offer when the phase is changed from amorphous (AM) to crystalline (CR), or vice versa. However, unwanted PCM infrared absorption is an issue, and for the prior-art GeSbTe (GST) and GeSbSeTe (GSST) PCMs, there is an infrared extinction problem in the crystalline phase. Prior work on GST and GSST revealed a large absorption at the 1550-nm telecom band. By contrast, recent research on the $Sb_2Se_2$ and $Sb_2S_3$ PCMs has attracted considerable interest because their extinction coefficients $k$ are extremely low at telecoms in both CR and AM [1]-[5]. We shall consider $Sb_2Se_3$ embedded in a GeOI strip waveguide (or guides) where the fundamental $TE_0$ mode is propagating and where the mode effective index is changed during phase change by a thin-film graphene micro-heater [6] that thermally contacts the Ge guiding region. If we look for a moment at the index difference $\delta n$ between the bulk AM and CR indices of $Sb_2Se_3$ at $\lambda \sim 2.5$ μm, we find that $\delta n \sim 0.5$, which is a smaller change than exhibited by GST and GSST. However, we have found that $\delta n$ of 0.5 produces highly significant switching performances and is still "formidable."

To our knowledge, this is the first analysis of PCMs in GeOI. Regarding the SOI platform, we note that the present DC and MZI approaches apply immediately to SOI with $Sb_2Se_3$ filling the slot in between Si strip waveguides or filling the slot within a Si strip. And that approach is valid over 1.55 μm to 8 μm (where silicon begins to cut off) because of low loss in $Sb_2Se_3$.

## 2.2 2 x 2 slot/pcm directional-coupler switch

In previous DC-switching studies, PCMs have been used as top cladding upon the central waveguides of two, three-waveguide (3W) and four-waveguide (4W) directional coupler [7]-[17]. The new aspect here is filling the lateral coupling zone between two waveguides (2W) with PCM to form a higher-performance device. Our view is that the resulting 2W switches are more compact, simpler, and more effective than the 3W- and 4W-DC 2×2s.

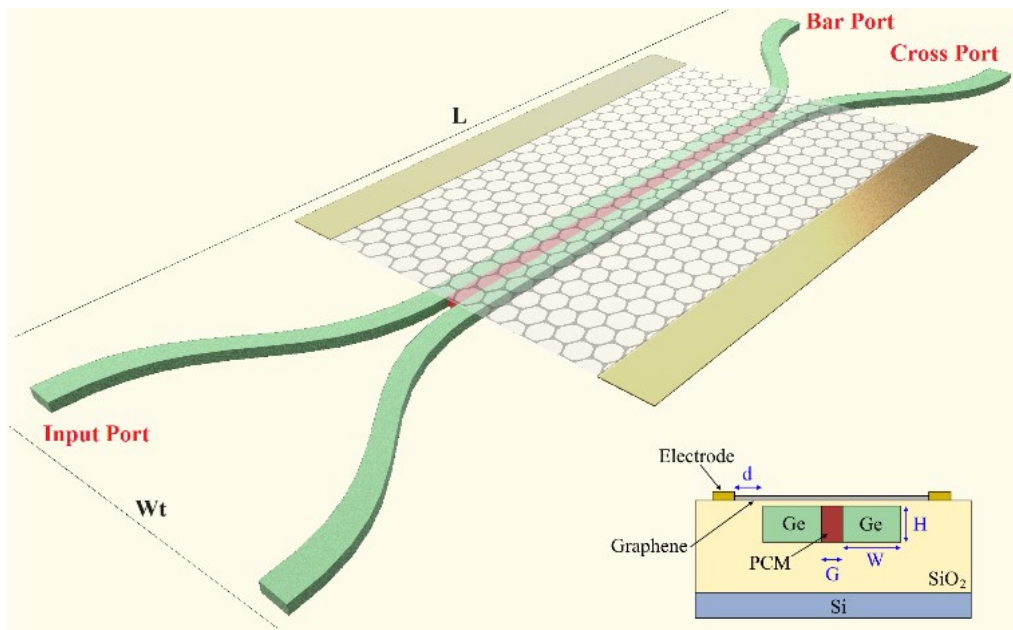

**Fig. 1**. *2 x 2 crossbar switch based upon $Sb_2Se_3$ embedded in the directional-coupler slot (one of two infrared inputs is shown). The graphene micro-heater is deposited on the planarized (with added oxide) strip-waveguides-on SOI surface. The inset shows the cross-section view in the coupling zone.*

Figure 1 illustrates the 2×2 EO crossbar switch actualized by $Sb_2Se_3$ embedded in the slot of the GeOI directional coupler. Independent optical signals are launched at the DC input ports. Each signal is polarized as the $TE_0$ fundamental mode, while the coupling zone includes some two-mode interference [3]. When $Sb_2Se_3$ is AM, the light will output at the Bar Port. After $Sb_2Se_3$ is switched to CR, the coupling coefficient in the coupling region will be modified accordingly due to refractive index change of $Sb_2S_3$ so that the light exits the Cross Port.

In addition, a very thin 20 nm oxide film is added atop the guides to give electrical isolation from the thin-film graphene resistive Joule heater that is deposited upon the active PCM and the oxide prevents PCM damage. The graphene film is etched into the form of a very localized heater with $V^+$ and Ground contacts. In summary, the GeOI waveguided 2×2 structure is planarized before graphene formation.

## 2.3  2 x 2 MZI Slot-waveguide Switch

The next proposed 2×2 GeOI crossbar structure is the compact MZI EO crossbar switch that is illustrated in Fig. 2, where each Ge-strip $W \times H$ waveguide arm contains a narrow slot of width $g$ and height $H$ that is filled with $Sb_2Se_3$.

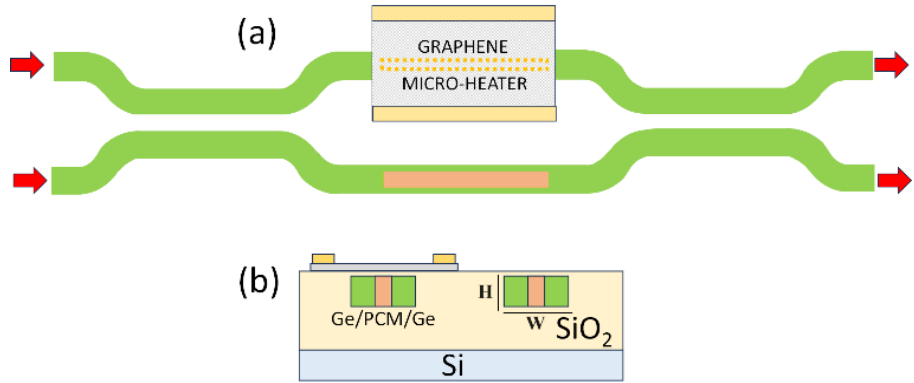

**Fig. 2**. *(a) 2 x 2 MZI crossbar switch having $Sb_2Se_3$ embedded in the slotted waveguides used for MZI arms. A localized graphene micro-heater is deposited on one arm. (b) Cross section view at the midline of the MZI showing the planarized surface for the entire device.*

The two passive 3-dB couplers used in Fig. 2 are consistent with Fig. 1 because they have the same S-bend waveguides of Fig. 1. In particular, our Fig. 2 simulations give an S-bend length of 7 μm, an S-bend height of 2.5 μm, a coupler gap of 100 nm, and a coupling zone length of 7.54 μm.

For the EO function, the graphene micro-heater is added to one MZI arm as illustrated in Fig. 2 and the cross-section indicates that we are utilizing the same planarized-$SiO_2$ approach to the strip waveguide MZI as was done for the Fig.1 DC. The input and output waveguides are the 600 nm x 400 nm strips. The MZI requirement is that an electro-optical phase shift (EOPS) of π radians, a retardation of the optical wave, must be attained in one arm of the interferometer to induce Cross to Bar. The Fig. 2 method is to deploy the slotted-Ge as the EOPS that shifts the $TE_0$ phase when the PCM slot is changed from CR to AM, a shift due to change in effective index of the waveguide. The $Sb_2Se_3$ slot is placed in both arms to equalize the Cross and Bar insertion losses (ILs) in the CR phase, and only one arm is "AM active."

Taking the $Sb_2Se_3$ slot width $g$ as 100 nm, and taking $\lambda$ = 2500 nm, we then simulated the MZI operation in order to find the slot length $L_{PCM}$ that provides π shift during CR to AM. The result is that $L_{PCM}$ = 7.84 μm is required for the complete crossbar switching. As a result, the area or "footprint" of the 2×2 MZI is 5 μm x 51 μm. The IL (Cross, CR) and IL (Bar, AM) are both estimated as -0.02 dB and bandwidths of 20 nm and 64 nm are recorded at crosstalk (CT) levels of -30 dB and – 20 dB, respectively.

If we extend both output waveguides by 9 μm in Fig. 2, then we can form a 7.8 μm slot in one of the lengthened arms, a slot filled with $Sb_2Se_3$. Then, by adding a second graphene controller, we can create a second independently addressed EOPS in that arm, giving an extended MZI, 5 μm x 60 μm in area, which is a highly competent, doubly addressed device that can be employed in meshes [18] and in "linear transformation devices" for optical neural networks [19]-[20].

## 2.4  Micro-Ring Resonator devices

Compact and capable resonant EO devices could be built using the same GeOI PCM technology. The favorable results found for the PCM-slotted Ge strip imply that a slot of length 1 or 2 μm will be effective as a "resonance shifter" if employed in the circular waveguide (or the racetrack waveguide) of a GeOI-strip micro-ring resonator (MRR) Such an MRR could also use the partial crystallization described below. Having a microns-scale graphene heater over the MRR slot could easily shift the resonance wavelength, thereby giving a 1×1 EO modulator or 2×2 EO switching, depending upon whether one or two bus waveguides are side coupled to the MRR. In addition, it is known that PCMs are promising candidates for realizing also zero-static power-consumption photonic memories [21], due to their reversible amorphous-crystalline phase transition, and their exceptional long-term, self-sustaining capability.

Considering the photonic integrated circuit as a whole, the entire group of PCM devices can be fabricated using an on-circuit photomask formed prior to deposition of PCM on the circuit. Small openings in the mask direct PCM into the various slots and DC coupling gaps, although complete PCM-filling of slots can be difficult if the slot width is much smaller than the slot height. After PCM deposition, all waveguide surfaces are to be planarized.

## 3. Results

### 3.1 2 x 2 DC Simulations

The goal of this section is to present design rules for the broadband DC switch. In the coupling section, the slot width is $G$ and the PCM active length ($L_{PCM}$) is designed to be equal to the coupling length ($L_c$). In the Fig. 1 inset, $W \times H$ represents the size of the GeOI waveguides. In the following analysis, we include the wavelength dispersion of material complex index by considering the Sellmeier equations for both germanium and silicon dioxide. For the PCM, the wavelength dependence of the real part of the refractive index ($n$) and the extinction coefficient ($k$) of the $Sb_2Se_3$ are taken into account, according to the measurements presented in [1]. However, in the non-absorbing region ($k \cong 0$ for $\lambda > 1600$ nm), the refractive index $n(CR)$ and $n(AM)$ could be fitted using the normal Cauchy's dispersion formula (adopted in our simulations) $n = n_0 + a/\lambda^2 + b/\lambda^4$, where $n_0$, $a$ and $b$ are constants given in [22].

We performed parametric simulations based on three-dimensional finite-difference time domain method, to evaluate the IL and CT metrics. To find design trends, several parametric simulations have been carried out on the Fig. 1 switch. The PCM slot widths $G$ of 100 nm, 130 nm and 160 nm were examined, using GeOI waveguide cross section of $W \times H$ = 600 nm × 400 nm. Figure 3 presents the three results for infrared transmission (output in dB) as a function of DC coupling length. Looking for excellent IL metrics in both Cross and Bar, an examination of Fig. 3 reveals the best trade-off as being $G$ = 130 nm and $L_c$ = 17 μm. Using that choice, we then calculate both the IL and CT as a function of wavelength, presented in Fig. 4. In that Figure, there are several possible $\lambda$-operating points, and here we shall choose $\lambda_0$ = 2494 nm to evaluate metrics.

We then find IL(Cross, CR) = 0.04 dB, IL(Bar, AM) = 0.22 dB, CT(Cross, AM) = -17 dB and CT(Bar, CR) = -34 dB, Defining this switch bandwidth as the wavelength range over which IL is less than 0.3 dB, we find BW = 30 nm, extending from 2471 nm to 2501 nm. Within this 30 nm, it results CT < -15 dB.

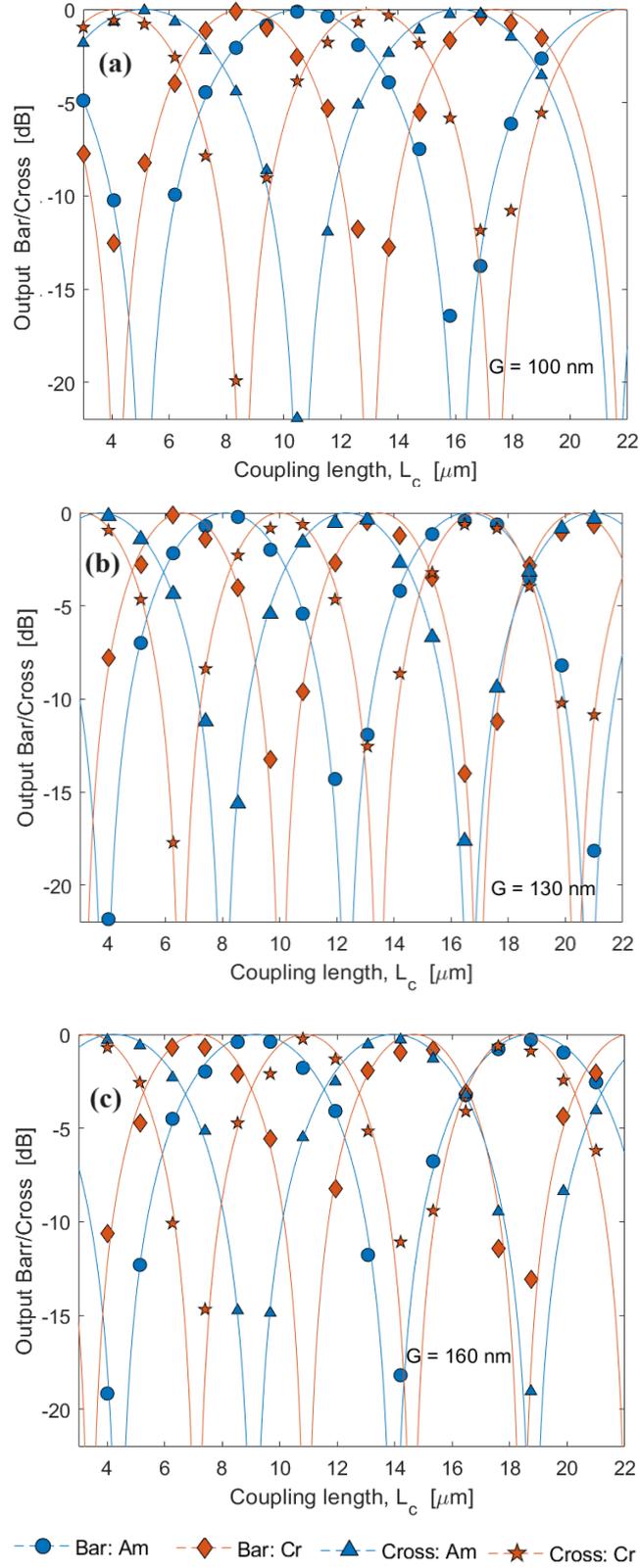

*Fig. 3*. *Output power at Bar and Cross Ports as a function of the coupling length, assuming λ= 2500 nm. (a) G=100 nm; (b) G=130 nm; (c) G=160 nm. Markers represent the 3D FDTD results and solid lines the curve fitting.*

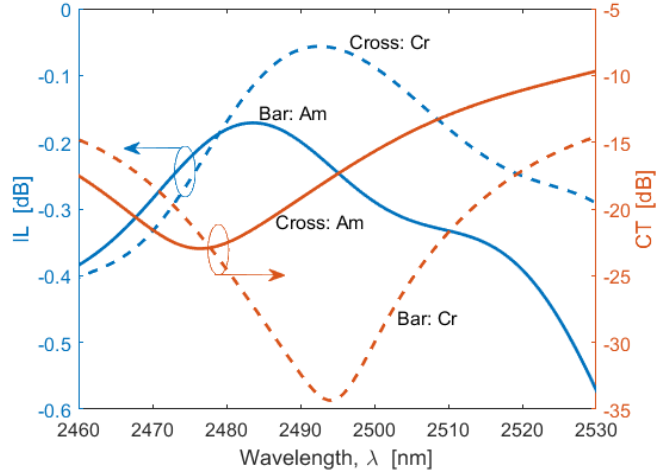

**Fig. 4.** *Insertion loss and Crosstalk at the Bar and Cross states as a function of the wavelength, for G=130 nm, and $L_C$ = 17 μm.*

Next, we investigated guided-mode intensity patterns of this optimized Fig. 2 DC with the results presented in Fig. 5. Here the excellent IL and DC are well visualized.

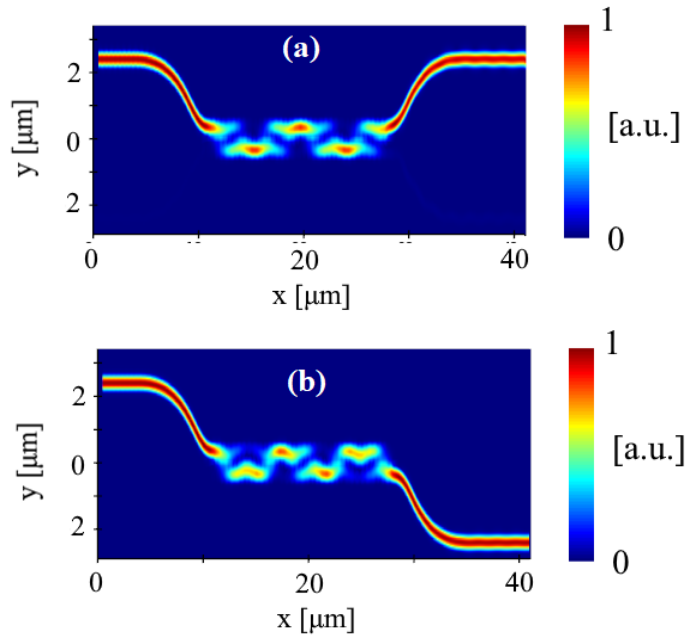

**Fig. 5.** *3D simulations of the 2×2 optical switch based on the $Sb_2Se_3$-slotted directional coupler. (a) Bar state (Am phase); (b) Cross state (Cr phase).*

It is interesting to make a "structural" comparison of our GeOI DC with the theoretical modelling-and-simulation of Song *et al*. [3] who proposed an SOI directional coupler with $Sb_2S_3$ embedded in the gap, and with 1.55 μm wavelength of operation. They proposed silicon rib waveguides for the coupler with *p+* and *n+* doping of side ribs to form a forward-biased *pin* diode in which electric current is passed through the $Si/Sb_2S_3/Si$ combination to provide Joule heating of the PCM. Such direct electrical heating of PCM is not optimum, and that our thin-film graphene-heater approach offering PCM temperature rise by thermal conduction is more advisable. This is confirmed in [23], where a comparison with *pin*, ITO, and graphene heaters has been performed. From the analysis, the graphene heater records excellent heating performance. In particular, graphene heaters show the highest switching speed thanks to the ultra-low heat capacity and high in-plane thermal conductivity of graphene, the ITO heaters exhibit a very large temperature gradient near the PCM because of the low thermal conductivity of ITO, and finally the *pin* heaters have severe thermal diffusion because of the high thermal conductivity of silicon. Also, from the standpoint of mode confinement and device networking, the rib waveguide method is less desirable than our strip waveguide approach. Indeed, as outlined in [24], one of the main limitations in reducing the footprint and cost of photonic integrated circuits is the minimum bending radius of the waveguides. It depends strongly on the size and lateral refractive index contrast of the waveguide. In this sense, the rib waveguide has low lateral index contrast which enables single-mode operation, but also severely limits the minimum bending radius [24].

Finally, we note that these 2×2s (Fig. 1) are versatile and effective "building blocks" for the construction of LSI networks, such as broadband N x N EO matrix switches. For example, Fig. 6 presents the quite-compact 8 x 8 nonblocking Benes matrix switch that uses only 20 crossbars and 16 intersections.

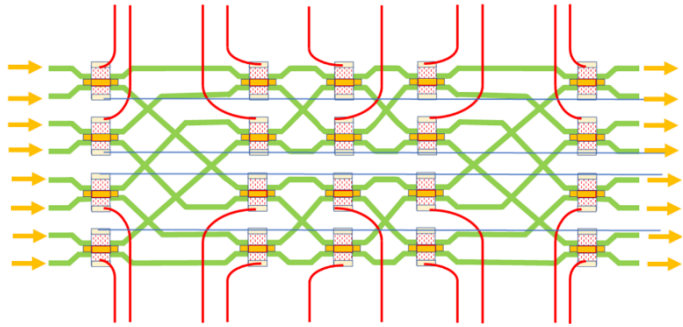

**Fig. 6.** *8x8 matrix switch based on the Benes architecture. The very compact directional-coupler PCM-slot crossbar switches are illustrated.*

## *3.2 Multi-Level Switching*

Returning to Fig. 2, partial amorphization is used to achieve stable, multiple levels. Starting from CR, the PCM is heated to a temperature higher than its crystallization temperature ($T_c$). The level is written by heating Sb$_2$Se$_3$ at a temperature $T_c < T \leq T_{am}$, by applying different voltage amplitudes. Consequently, the PCM slot is partially amorphized as induced by the thermal gradient along $z$, thus generating multi-levels in the optical output power (Bar and Cross ports). In particular, when the initial CR Sb$_2$Se$_3$ is amorphized by a 7.75 V pulse of 500 ns duration applied to the graphene heater, (see thermal simulations), a change in the effective refractive index ($\Delta n_{eff} = n_{eff}(CR) - n_{eff}(AM)$) of 0.16 is recorded, for the fundamental TE modes. As a result, a Sb$_2$Se$_3$/Ge slot length, $L_{PCM} \cong 7.84$ μm is required to induce the Cross/Bar switching. Moreover. in our simulations we have assumed the Sb$_2$Se$_3$ slot width ($g$) of 100 nm. However, here, we systematically characterize the effect of the amplitude of voltage pulses upon the multilevel switching response of photonic memories. The Sb$_2$Se$_3$ slot is gradually amorphized to generate 41 levels in the PCM by applying an electric pulse with a duration ($t_p$) of 500 ns and voltage amplitudes in the range 6.39 V to 7.1 V (see the Thermal Simulations section). To realize a programmable optical convolutional neural network, we calculate the power contrast $\tau$ by evaluating the power of the upper (Bar) and lower (Cross) output (see Fig. 2 (a)). According to [20], the power contrast $\tau$ is given by:

$$\tau = \frac{T_{Bar} - T_{Cross}}{T_{Bar} + T_{Cross}} \qquad (1)$$

where $T_{Bar}$ and $T_{Cross}$ are the transmission efficiencies of the upper and lower output channels at 2500 nm, respectively. Normalized to the propagation losses, the parameter $\tau$ ranges from -1 to 1, depending on the degree of the partial amorphization which, in turn, induces the Sb₂Se₃ refractive index to change in the range $n(CR) \div n(AM)$ (see the thermal simulations section). In this context, Fig. 7 shows the multilevel programmability of the power contrast, $\tau$, as a function of the applied voltage amplitude, where a voltage amplitude resolution of 15 mV is assumed [21].

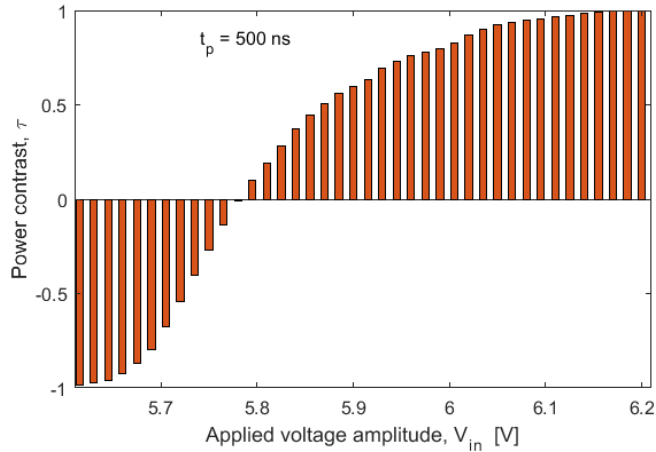

*Fig. 7. Power contrast, τ, for the MZI device as a function of the applied voltage amplitude. In the simulations the pulse duration is assumed as $t_p$=500 ns.*

The partial-amorphization approach to multi-level switching just quantified applies immediately to the Fig. 1 directional coupler. With the DC initially in the CR state, applied-voltage amplitudes similar to those listed above, will produce the quasi-continuous staircase switching in a manner similar to that in Fig. 7. Therefore, that DC offers nonvolatile almost-continuous states between Cross and Bar which can be described as a "tunable beam-dividing". For that reason, an interconnected network of these multi-level 2×2 DCs fulfills the application presented in Fig. 5(a) of Chen *et al*. [25], where they show this active photonic circuit providing nonvolatile forward matrix-vector multiplication in a weighted, deep neural network.

The capability of the Fig. 1 device can be expanded by extending both output waveguides by about 9 μm, and by then forming a slot with $G \times L$ = 130 nm x 7800 nm, after which Sb₂Se3 is embedded therein, and with a graphene electro-thermal controller then added atop that slot. The resulting device is shown in Fig. 8, where we show the two corresponding micro heaters that give multilevel switching in the coupling zone and in the external arm where "tunable phase shifting" is attained.

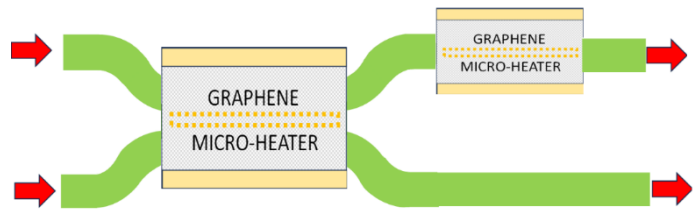

*Fig 8. 2 x 2 tunable cross-to-bar splitter and linked tunable phase shifter, with independent graphene controllers and 5 μm x 40 μm overall footprint.*

The new double-addressed device has a compact footprint of 5 μm x 40 μm and its quasi-continuous signal dividing and signal-phase shifting have an immediate application to nonvolatile reconfigurable gate arrays, illustrated in Fig. 6(b) of Chen *et al*. [25], where a rectangular mesh is shown to provide several varieties of integrated optical filters and delay lines. As a final comment, it is also quite feasible to create a Fig. 2 MZI having tuned beam-splitting and attached waveguided phase shifting, but the device area in that case is 5 μm x 60 μm.

## 3.3 Thermal Simulations of DC Switch

We investigated the thermal aspects of Fig. 1 and modeled the amorphization/crystallization mechanism using the Finite Element Method (FEM) models. We targeted the DC having 130 nm gap width and 17 μm length filled with Sb$_2$Se$_3$ and used GeOI strip waveguides having a cross section $W \times H$ = 600 nm × 400 nm. A 20 nm-thick SiO$_2$ film is assumed on the top to avoid oxidization of Sb$_2$Se$_3$. Moreover, a monolayer of graphene having a sheet resistance ranging from $R_s$=800 Ω to $R_s$=2000 Ω [26] is assumed on the planarized surface of the crossbar switch device. The graphene electrical conductivity is evaluated as $\sigma = 1/d_{gr}R_s$, where $d_{gr}$ is the graphene thickness. Our investigations indicate that here 8.7 V pulse of $t_p$=500 ns duration applied to the graphene resistor is required to induce the complete amorphization ($T \geq T_m \sim$ 884 K), as a result of the different material distribution.

To consider the time evolution of the temperature upon pulse excitation, we performed 3D FEM simulations where the heat-transfer-in-solids model is coupled together with the electric currents model. The time-dependent study is thus adopted to solve the coupled thermal and electrical equations under the rectangular pulse voltage excitation. The material properties used in the simulations are summarized in Table 1.

It is worth outlining that the temperature field presents a gradient along the directions y and z (Fig. 11 below). Therefore, the sections where the minimum heating is observed are located at the extremes of the PCM slot. At these sections, the temperature gradient along z gives a minimum and a maximum temperature ($T_{min}$, $T_{max}$) on the bottom and on the top of the Sb$_2$Se$_3$ slot, respectively.

*Table 1*

*Physical Parameters of Thermal Simulation*

| Parameters | Materials | | | |
|---|---|---|---|---|
| | Silicon dioxide | Germanium | AM-Sb$_2$Se$_3$ | CR- Sb$_2$Se$_3$ |
| Density [kg/m$^3$] | 2203 | 5323 | 5843 [27] | 6492 [27] |
| Thermal conductivity [W/mK] | 1.38 | 60 | ~0.2 [27] | ~0.24 [27] |
| Heat capacity at constant pression [J/kgK] | 746 | 310 | 507 [27] | 574 [27] |
| Electrical conductivity [S/m] | | | 4.7e$^{-7}$ | 4.7e$^{-7}$ |

Figure 9 shows the minimum and maximum temperatures ($T_{min}$, $T_{max}$) inside the PCM layer as a time function under rectangular voltage pulse excitation of 8.7V/500 ns, indicating, clearly that all of the Sb$_2$Se$_3$ slot is amorphized. In the simulations, we have assumed the metal electrode span ($d$ in the inset of Fig. 1) of 1500 nm. The quenching rate $R_q$ of the system has to be larger than the critical cooling rate ($R_c$) to avoid recrystallization. Due to the lack of the value in literature for Sb$_2$Se$_3$, we assume here, as a reference, the value for the GST-225, $R_c$=1K/ns. However, by comparing the times for the crystallization process (from ns to μs for GST-225 and from μs to ms for Sb$_2$Se$_3$ [1]), a $R_c$ smaller 1K/ns is expected for Sb$_2$Se$_3$. According to [3], the quenching rate ($R_q$) is calculated as $R_q = (T_m - T_c)/t_q$, where $t_q$ is the quenching time defined as the duration of temperature drop from the melting temperature $T_m$ to the crystallization temperature $T_c$. Figure 9 shows the transient temperature reaches a temperature maximum of ~1002 K at 490 ns and drops from melting temperature to crystallization temperature in $t_q$=290ns, inducing $R_q$=1.42 K/ns, which prohibits recrystallization.

However, the distance $d$ can be further reduced to lower the drive voltage, as shown in Fig. 10 where the temperature is plotted as a function of thickness in the z direction in the PCM layer ($z$ =0 and $z$ =400 nm correspond to the bottom and the top of the Sb$_2$Se$_3$ slot, respectively).

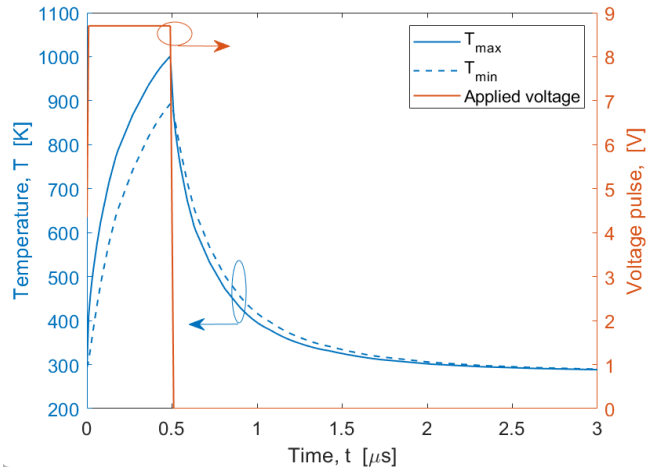

**Fig. 9.** *Maximum and minimum temperature in the PCM layer as a function of time for rectangular pulse heating.*

The curves of Fig. 10 indicate an increasing of the temperature above 1000 K for *d*=1300 nm, indicating the possibility of reducing the voltage amplitude below the 8.7 V needed for the complete amorphization at *d*=1500 nm. A narrower distance, *d*, could result in a higher propagation loss (induced by the metal electrodes). Thus, in a conservative way, *d* is chosen to be 1500 nm in our design.

To induce re-crystallization, the $Sb_2Se_3$ needs to be heated above the crystallization temperature ($T_c$~ 473 K) but below the melting point $T_m$. Although the operative temperature is lower than the melting temperature, this process is demonstrated to require a larger amount of time. According to the Table 2 of Ref. [1], the times for the crystallization and the times for amorphization range from few μs-to-ms and few ps-to-ns, respectively. For an example, our simulations indicate that $T_c < T < T_m$ can be achieved for a voltage pulse of 4.5V amplitude, and of duration 50 μs, for $R_s$=800 Ω.

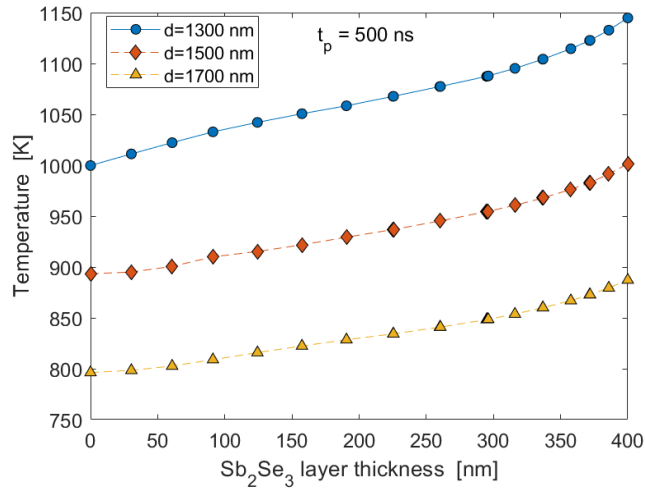

**Fig. 10.** *Temperature distribution (at $t_p$=500 ns) along the PCM layer, for a rectangular 8.7 V heating pulse.*

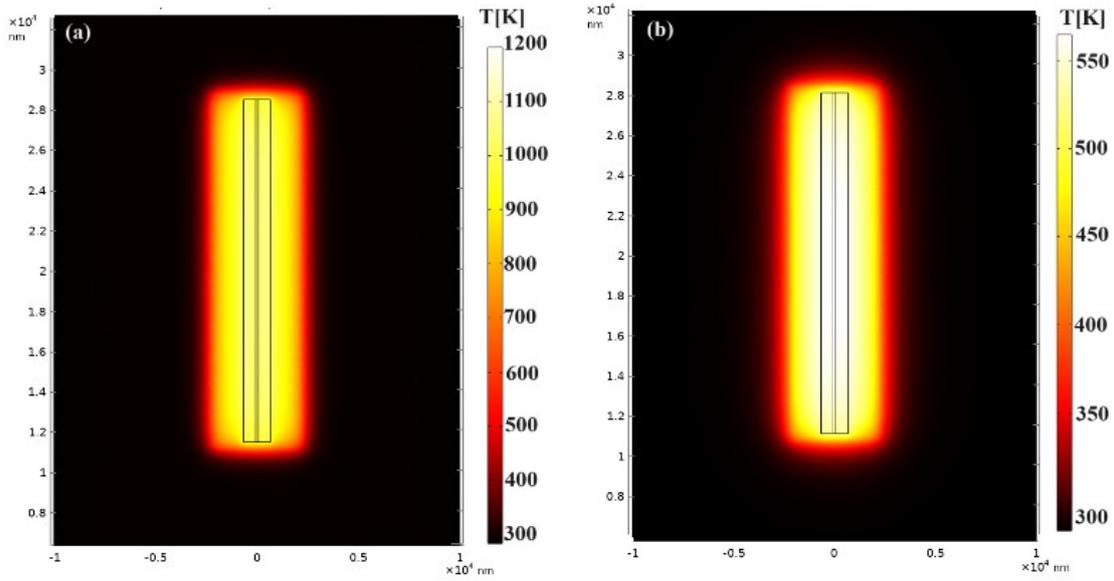

**Fig. 11.** *Temperature distribution in the longitudinal cross section plane, for the 2×2 optical switch based upon the slotted directional coupler architecture. (a) Amorphization; (b) Crystallization. Close-up view of the coupler region is shown.*

The temperature distributions along the longitudinal cross section plane, for the processes of amorphization and crystallization, are shown in Fig. 11 (a) and (b), respectively, where a zoomed-in view of the coupling region ($L_c$ = 17 µm) has been considered in order to highlight the temperature gradient. The simulations of Fig. 11 indicate that complete amorphization (crystallization) can be induced over the entire $L_c$ = 17 µm for the considered voltage pulses. In addition, we show in Fig. 12 the temperature distribution in the cross section of the 2×2 optical switch based upon the slotted directional coupler. The plots indicate clearly that the amorphization (crystallization) process starts to take place on the top of the $Sb_2Se_3$ layer, where the temperature first reaches the melting (crystallization) temperature value.

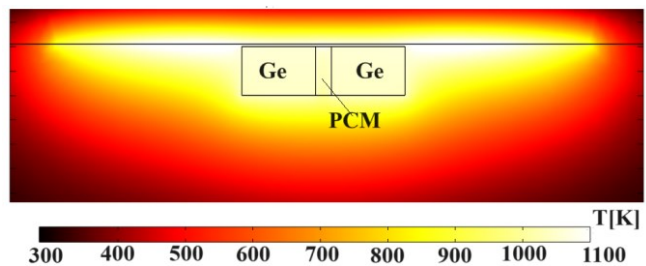

**Fig. 12.** *Temperature distribution in the cross section, for the 2×2 optical switch based upon the slotted directional-coupler architecture. (a) Amorphization; (b) Crystallization. Close-up view of the coupler region is shown.*

## 3.4 Thermal Aspects of Multilevel MZI

Similar investigations have been performed for the MZI architecture of Fig. 2 (a). However, here we focus on the effect of partial amorphization needed to realize multilevel photonic memories or for quasi-continuous switching between cross and bar. In this context, the PCM is all initialized to the crystalline state. This is achieved by heating the PCM up via the graphene heater to a temperature higher than its crystallization temperature ($T_c$) and holding that for a period of time (see previous section). The phase-change degree is then defined as the ratio of the amorphization depth to the total film thickness [28]. In the experimental investigations proposed in [28], the phase-change degree is controlled by a combination of the pulse energy and the number of pulses in multi-pulse femtosecond laser irradiation for amorphization. However, here we adopt the experimental approach of [21], where only the pulse energy is controlled. The multilevel amorphization is thus realized by applying an electric pulse with a fixed duration and various voltage amplitudes not exceeding the threshold value for the complete amorphization. We have focused on the active arm of the MZI device, where the GeOI cross section is $W \times H =$ 600 nm $\times$ 400 nm, $L_{PCM}$=7.84 μm, and $g$ = 100 nm.

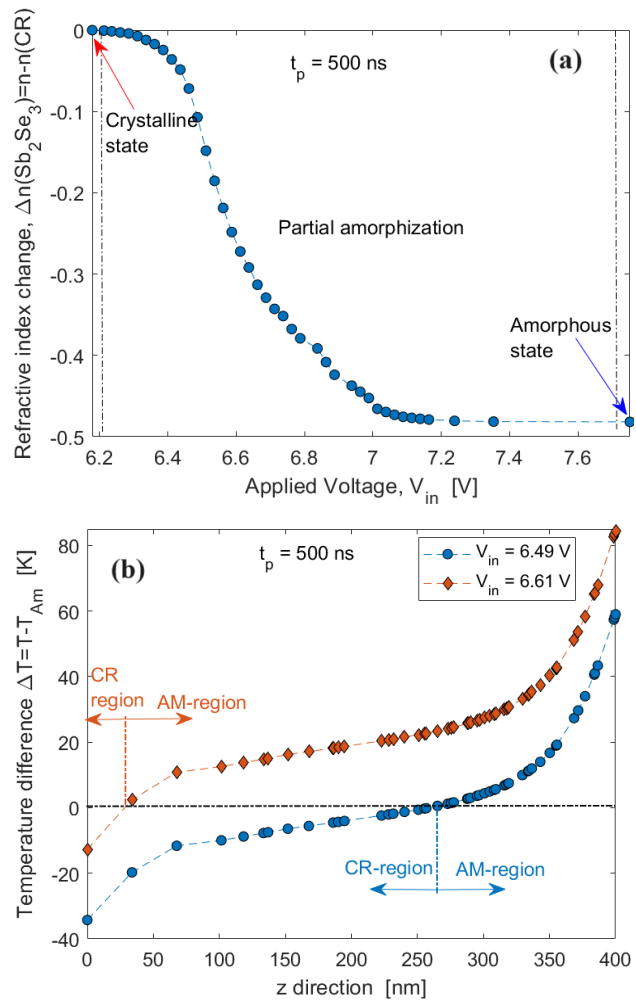

*Fig. 13.* (a) Average refractive index change of the Sb$_2$Se$_3$ as a function of the applied voltage amplitude (at $t_p$=500 ns); (b)Temperature distribution (at $t_p$=500 ns) vertically along the PCM layer, assuming two applied voltage pulses of 6.49 V and 6.61 V.

The results are plotted in Figs. 13 (a) and (b). In particular, Fig. 13 (a) shows the average refractive index change ($\Delta n(Sb_2Se_3) = n(Sb_2Se_3) - n(CR)$)) in the Sb$_2$Se$_3$ slot as a function of the amplitude voltage, assuming the pulse duration $t_p$=500 ns. Here, $n(CR)$ and $n(Sb_2Se_3)$ represent the values of the refractive index under the conditions of complete crystallization and partial amorphization, respectively. The plot indicates that the total and partial amorphization occur for $V_{in} \geq 7.75$ V and 6.18 V$< V_{in} <$ 7.75 V, respectively. The effect of the gradual amorphization is well evidenced in Fig. 13 (b) where the temperature inside the Sb$_2$Se$_3$ slot is shown as a function of the $z$ direction, for two values of the applied voltage amplitude. For $V_{in}$= 6.49 V and 6.61 V, the thickness of Sb$_2$Se$_3$ amorphized (evaluated from the top) changes from

135 nm to 366 nm, respectively. When $V_{in}$=6.39 V, the $Sb_2Se_3$ slot is predominantly in the crystalline phase and the MZI effectively transmits the $TE_0$ mode light in the Cross channel, with a power ratio of 98.12% (see Fig. 7). Conversely for $V_{in}$=7.1 V the $Sb_2Se_3$ layer is predominantly in an amorphous state, and the $TE_0$ mode is recorded at the Bar channel with a power ratio of 99.82% (see Fig. 7). It is worth noting that even if it is possible to consider applied voltage amplitude $V_{in}$ <6.39 V (total crystallization) and 7.1 V≤ $V_{in}$ ≤ 7.75 V, these values are not shown in Fig. 7 since the levels should be not distinguishable.

At this step, some comments on the energy consumption are worth making. For example, the total amorphization process requires an energy consumption ranging from 192 nJ (83.9 nJ) to 77 nJ (33.56 nJ) for $R_s$=800 Ω and $R_s$=2000 Ω, respectively and for DC switch (MZI). Although the cross section of our devices is larger than that of SOI switches (longer wavelength operation), we outline that the calculated energy consumptions are lower than the values proposed in [15], and 3 or 4 times larger than the results listed in [26]. However, the device in [26] is a simple phase shifter having a total PCM length of 4.93 μm against the values of 17 μm or 7.84 μm used here and then resulting in an increased total graphene resistance. We think that further optimizations of the graphene configuration oriented to increase the contact resistance with the electrode will help to reduce the energy consumption. Moreover, this optimization would allow an endurance of over 1,000 cycles, according to the experimental measurements of [26].

## 4. Conclusion

The photonics platform assumed in this work is the foundry-compatible and CMOS-compatible GeOI platform operating in the MWIR band at 2.5 μm. A monolayer resistive-graphene micro-heater provides electro-optical control of PCM embedded in Ge waveguides. In this paper, silicon-based, manufacturable on-chip nonvolatile broadband 2×2 crossbar switches are designed and analyzed theoretically. One switch consists of a two-waveguide coupler where ultra-low-loss $Sb_2Se_3$ is embedded in the gap between coupled GeOI strip waveguides, and where the Bar or Cross state is induced by setting PCM in its AM or CR phase, respectively. The graphene is in thermal contact with the top of the slot. The switch footprint is 5 μm x 31 μm. The second switch is an MZI wherein both Ge strip arms have $Sb_2Se_3$ embedded in a 7.8-μm-length slot, and where the device footprint is 5 μm x 51 μm. The passive GeOI waveguides have a cross section of $W \times H$ = 600 nm × 400 nm.

Rectangular voltage pulses having a 500-ns duration are applied to the graphene electrodes. Multilevel switching and photonic memories can be based upon DC and MZI and MRR. Our simulations indicate the feasibility of "storage" at many levels, such as 41, by carefully controlling the applied voltage amplitude within 6.18 V< $V_{in}$ < 7.75 V.

## Acknowledgements

R.S. wishes to thank the U.S. Air Force Office of Scientific Research for their sponsorship on grant FA9550-21-1-0347. This work was partially supported by the European Union under the Italian National Recovery and Resilience Plan (NRRP) of NextGenerationEU, partnership on "Telecommunications of the Future" (PE00000001 - program "RESTART").

## References

[1] D. Tripathi, H. S. Vyas, S.Kumar, S. S Panda, R. Hegde, "Recent developments in Chalcogenide phase change material-based nanophotonics," *Nanotechnology,* **34,** 502001, 1-29, (2023).

[2] H. J. Kim, M. Julian, C. Williams, D. Bombara, J. Hu, T. Gu, K. Aryana, G. Sauti, W. Humphreys, "Versatile spaceborne photonics with chalcogenide phase-change materials," *arXiv preprint*, arXiv:2310.19131

[3] C. Song, Y. Gao, G. Wang, Y. Chen, P. Xu, C. Gu, Y. Shi, X. Shen, "Compact nonvolatile 2×2 photonic switch based on two-mode interference," *Opt. Express,* **30**(17), 30430-30440, (2022).

[4] M. Delaney, I. Zeimpekis, D. Lawson, D. W. Hewak, O. L. Muskens, "A New Family of Ultralow Loss Reversible Phase-Change Materials for Photonic Integrated Circuits: $Sb_2S_3$ and $Sb_2Se_3$," *Adv. Funct. Mater.*, **30**, 2002447, (2020).

[5] J. Faneca, I. Zeimpekis, S. T. Ilie, T. D. Bucio, K. Grabska, D. W. Hewak, F. Y. Gardes, "Towards low loss non-volatile phase change materials in mid index waveguides, *Neuromorph. Comput. Eng.* **1,** 014004, (2021).

[6] J. Faneca, S. Meyer, F. Y. Gardes, D. N. Chigrin, "Graphene microheater for phase change chalcogenides based integrated photonic components," *Optical Materials Express,* **12**, 5, 1991-2002, (2022).

[7] H. Liang, R. Soref, J. Mu, X. Li, and W.-P. Huang, "Electro-optical phase-change 2x2 switching using three- and four-waveguide directional couplers," *Appl. Opt*., vol. **54**, no. 19, pp. 5897-5902 (2015).

[8] Y. Zhang, J. Li, J. Chou, Z. Fang, A. Yadav, H. Lin, Q. Du, J. Michon, Z. Han, Y. Huang, H. Zheng, T.Gu, V. Liberman, K. Richardson, and J.Hu, "Broadband Transparent Optical Phase Change Materials," CLEO: Applications and Technology 2017, San Jose, California, United States, 14–19 May 2017, ISBN: 978-1-943580-27-9, jTh5C.4.

[9] P. Xu, J. Zheng, J. K. Doylend, and A. Majumdar, "Low-Loss and Broadband Nonvolatile Phase-Change Directional Coupler Switches," *ACS Photonics* **6**(2), 553–557 (2019).

[10] C. Rios, M. Stegmaier, Z. Cheng, N.Youngblood, C. D.Wright, W. H. P. Pernice, and H. Bhaskaran, "Controlled switching of phase-change materials by evanescent-field coupling in integrated photonics [Invited]," *Opt. Mat. Express*, **8**(9), 2455-2470 (2018).

[11] N. Ali, R. Kumar, "Mid-Infrared non-volatile Silicon photonic switches using $Ge_2Sb_2Te_5$ embedded in SOI waveguide," *Nanotechnology*, **31**, (11), (2019).

[12] J. Song, S. Ghosh, N. Dhingra, H. Zhang, L. Zhou, and B. M. A. Rahman, "Feasibility study of a Ge2Sb2Te5-clad silicon waveguide as a non-volatile optical on-off switch," *OSA Continuum*, **2**(1), 49-63 (2019).

[13] W. Jiang, "Nonvolatile and ultra-low-loss reconfigurable mode (de)multiplexer/switch using triple-waveguide coupler with $Ge_2Sb_2Se_4Te_1$ phase change material," *Scientific Reports,* **8**(1), 15946 (2018).

[14] W. Jiang, "Reconfigurable Mode (De)Multiplexer via 3D Triple-Waveguide Directional Coupler with Optical Phase Change Material," *IEEE J. Lightwave Technology*, **37**, 1000-1007, (2019).

[15] F. De Leonardis, R. Soref, V. M. N. Passaro, Y. Zhang, and J. Hu, "Broadband Electro-Optical Crossbar Switches Using Low-Loss Ge2Sb2Se4Te1 Phase Change Material," *IEEE J. Lightwave Technol*. **37**(13), 3183–3191 (2019).

[16] H. Liang, R. Soref, J. Mu, A. Majumdar, X. Li, and W. Huang, "Simulations of silicon-on-insulator channel-waveguide electro-optical 2×2 switches and 1×1 modulators using a $Ge_2Sb_2Te_5$ self-holding layer," *IEEE J.. Lightwave Technol*. **33**(9), 1805–1813 (2015).

[17] Y. Zhang, Q. Zhang, J. B. Chou, J. Li, C. Roberts, M. Kang, C. Gonçalves, R. Soref, C. Ríos, M. Shalaginov, K. Richardson, T. Gu, V. Liberman, J. Hu, "Designing nonvolatile integrated photonics with low-loss optical phase change materials," *SPIE Proceedings,* **11081**, 110811S (2019).

[18] F. Shokraneh, M. Sanadgol Nezami, O. Liboiron-Ladouceur, "Theoretical and Experimental Analysis of a 4×4 Reconfigurable MZI-Based Linear Optical Processor," *IEEE J.. Lightwave Technol*., **38**(6), 1258-1267, (2020).

[19] H. Zhang, M. Gu, X.D. Jiang, *et al.* "An optical neural chip for implementing complex-valued neural network," *Nat. Commun* . **12**, 457 (2021).

[20] H. Yuan, Z. Wang, Z. Peng, J. Wu, J.Yang, "Ultra-Compact and NonVolatile Nanophotonic Neural Networks," *Adv. Optical Mater.* 2300215 (1-10), (2023).

[21] M. Wei, J.Li, Z. Chen, B. Tang, Z. Jia, P. Zhang, K. Lei, K. Xu, J. Wu, C. Zhong, H. Ma, Y. Ye, J. Jian, C. Sun, R. Liu, Y. Sun, W. E. I. Sha, X. Hu, J. Yang, L. Li, H. Lin, "Electrically programmable phase-change photonic memory for optical neural networks with nanoseconds in situ training capability," *Advanced Photonics*, **5**(4), 046004, (2023).

[22] C. Chen, W. Li, Y. Zhou, C. Chen, M. Luo, X. Liu, K. Zeng, B. Yang, C. Zhang, J. Han, J. Tang, "Optical properties of amorphous and polycrystalline Sb2Se3 thin films prepared by thermal evaporation," *Appl. Phys. Lett.* **107**, 043905-1-5, (2015).

[23] J. Zheng, S. Zhu, P. Xu, S. Dunham, A. Majumdar, "Modeling Electrical Switching of Nonvolatile Phase-Change Integrated Nanophotonic Structures with Graphene Heaters," *ACS Appl. Mater. Interfaces,* **12**, 21827−21836, (2020).

[24] M. Cherchi, S. Ylinen, M. Harjanne, M. Kapulainen, and T. Aalto, "Dramatic size reduction of waveguide bends on a micron-scale silicon photonic platform," *Optics Express*, **21**(15), 17814-17823, (2013).

[25] R. Chen, Z. Feng, F. Miller, H. Rarick, J. Froch, A. Majumdar, "Opportunities and challenges for large-scale phase-change material integrated electro-photonics," *ACS Photonics,* vol **9**, 3181-3195 (2022).

[26] Z. Fang, R. Chen, J. Zheng, A. I. Khan, K. M. Neilson, S. J. Geiger, D. M. Callahan, M. G. Moebius, A. Saxena, M. E. Chen, C. Rios, J. Hu, E. P. A. Majumdar, "Ultra-low-energy programmable non-volatile silicon photonics based on phase-change materials with graphene heaters*," Nature Nanotechnology*, **17**, 842-848, (2022).

[27] H. Zhu, Y. Lu, L. Cai, "Wavelength-shift-free racetrack resonator, hybrided with phase change material for photonic in-memory computing," *Optics Express*, **31**(12), 18840-18850, (2023).

[28] K. Gao, K. Du, S. Tian, H. Wang, L. Zhang, Y. Guo, B. Luo, W. Zhang, Ting Mei, "Intermediate Phase-Change States with Improved Cycling Durability of $Sb_2S_3$ by Femtosecond Multi-Pulse Laser Irradiation", *Adv. Funct. Mater.,* 2103327 (1-7), (2021).